\documentclass[acus]{pac99}

\usepackage{epsfig}

\newcommand{\bit}{\begin{Itemize}}
\newcommand{\eit}{\end{Itemize}}

\begin{document}
\title{A CUPRONICKEL ROTATING BAND PION PRODUCTION
TARGET FOR MUON COLLIDERS$^1$}
\author{ B. J. King$^2$, S.S. Moser$^3$, R.J. Weggel,
BNL$^4$; N.V. Mokhov, Fermilab$^5$}
\maketitle

\begin{abstract}

A conceptual design is presented for a high power cupronickel pion
production target. It forms a circular band in a horizontal plane with
approximate dimensions of: 2.5 meters radius, 6 cm high and 0.6 cm thick.
The target is continuously rotated at 3 m/s to carry heat away from the
production region to a water cooling channel. Bunches of 16 GeV protons
with total energies of 270 kJ and repetition rates of 15 Hz are incident
tangentially to an arc of the target along the symmetry axis of a 20 Tesla
solenoidal magnetic capture channel.
The mechanical layout and cooling setup are described. Results are
presented from realistic MARS Monte Carlo computer simulations of the pion
yield and energy deposition in the target. ANSYS finite element calculations
are beginning to give predictions for the resultant shock heating stresses.

\end{abstract}

\footnotetext[1]{This work was performed under the auspices of
the U.S. Department of Energy under contract no. DE-AC02-98CH10886. }
\footnotetext[2]{Submitting author, email: bking@bnl.gov .}
\footnotetext[3]{on academic leave from Saint Joseph's College, Indiana.}
\footnotetext[4]{Brookhaven National Laboratory, P.O. Box 5000,
Upton, NY 11973-5000}
\footnotetext[5]{Fermi National Accelerator Laboratory, P.O. Box 500,
Batavia, IL 60510-0500}

\section{Introduction}
\label{sec-intro}

 High power pion production targets are required in current
scenarios~[1] for muon colliders. The pion secondaries
from protons on the target are captured in a solenoidal magnetic
channel and decay into the muon bunches needed for cooling,
acceleration and injection into the collider ring. Bunched proton
beams of several megawatts will be needed for the currently
specified muon currents~[1]: approximately
$6 \times 10^{20}$ muons of each sign at repetition rates
of 15 Hertz and in bunches of up to $4 \times 10^{12}$ muons
per bunch. This is an extrapolation from today's high power
targets in rate of target heating, shock stresses and integrated
radiation damage to the target.

 Because of the high beam power, liquid metal jet targets have been
the subject of much recent study and form the bulk
of a proposed experimental R \& D program of targetry studies that has
recently been submitted~[2] to the BNL AGS Division.
More conventional solid targets have several challenges. Along
with concerns about shock heating stresses and radiation damage, it
is challenging to design a cooling scenario consistent with both the 
large beam power and the small target cross sections that are needed
for high pion yields.

 This paper introduces the idea of a solid target in the form of a
band that addresses this cooling issue by rotating the band to carry
heat away from the targetry region and through a cooling channel.

\begin{figure}[t!] 
\centering
\epsfig{file=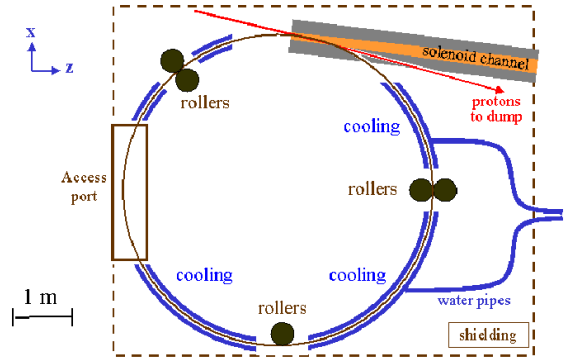, angle=270, width=3.0in}
\caption{A conceptual illustration of the targetry setup.}
\label{layout}
\end{figure}

\begin{figure}[t!] 
\centering
\epsfig{file=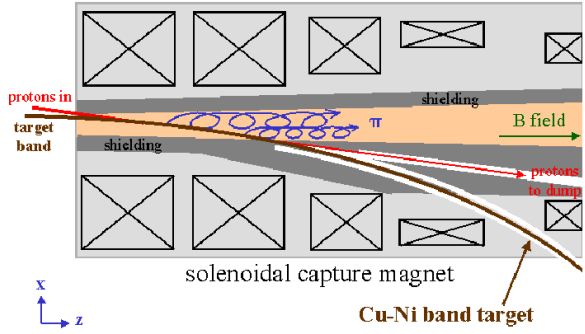, angle=270, width=3.0in}
\caption{A conceptual illustration of the target layout around the
pion production region.}
\label{closeup}
\end{figure}

\begin{figure}[t!] 
\centering
\epsfig{file=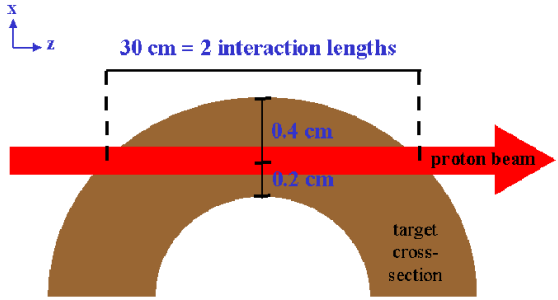, angle=270, width=3.0in}
\caption{The trajectory of the proton beam into the target band.}
\label{band}
\end{figure}

 Figures~\ref{layout} and~\ref{closeup} give {\em schematic} views
of the targetry setup we are considering and figure~\ref{band}
shows the trajectory of the proton beam into the target band.
It must be emphasized that details such as the rollers and cooling
setup are only shown schematically and no effort has been put into
their design.

The target band is enclosed in a 20 Tesla
solenoidal magnetic pion capture magnet whose general design
has previously been studied~[1]
by the Muon Collider Collaboration (MCC). The major design modification
specific to this particular geometry concerns the provision of entry
and exit ports for the target band.

 The high-power bunched proton beam
strikes the target band at a glancing angle and travels along inside the
target material for two nuclear interaction lengths before the curvature
of the band brings it again to an exit point at the outer edge of the band.
The beam is tilted at 150 milliradians to the
longitudinal axis of the solenoidal magnet;
MARS simulations described below show that this gives a larger
pion yield than a beam parallel to the solenoid.

\section{MARS Simulations of Target Heating and Pion Yield}

Full MARS~[3] tracking and showering Monte Carlo simulations were
conducted for a 16 GeV proton beam of
1$\times$10$^{14}$~ppp with a repetition rate of 15~Hz
on Ni band (R=250~cm, 6~cm height, 0.6~cm thickness) in a 20~T solenoid of
R$_a$=7.5~cm half-aperture.
Both untilted targets and targets tilted by $\alpha$=150~mrad were
studied and detailed 3-dimensional maps of energy deposition densities
were generated for input to the ANSYS stress analyses.

\begin{figure}[hbt]
\centering
\epsfig{file=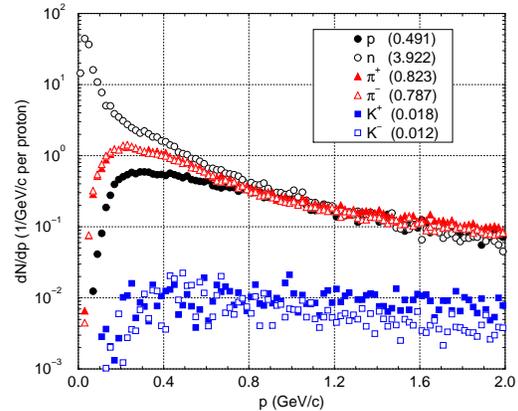, width=3.0in}
\caption{Momentum spectra of hadrons at L=90~cm and R$<$7.5~cm for a tilt
angle  $\alpha$=150~mrad. Integrated yield is shown in parentheses.} 
\label{phadron}
\end{figure}

\begin{figure}[hbt]
\centering
\epsfig{file=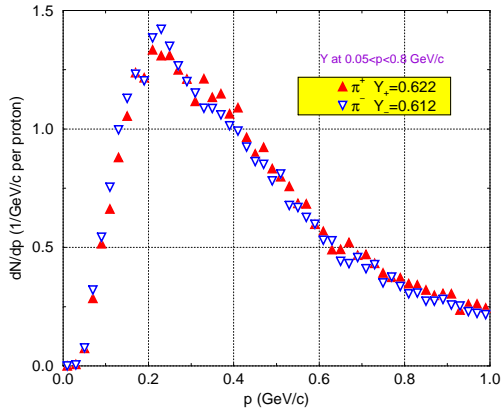, width=3.0in}
\caption{Pion momentum spectra at L=90~cm and R$<$7.5~cm
for a tilt angle $\alpha$=150~mrad.} 
\label{ppion}
\end{figure}

\begin{figure}[hbt]
\centering
\epsfig{file=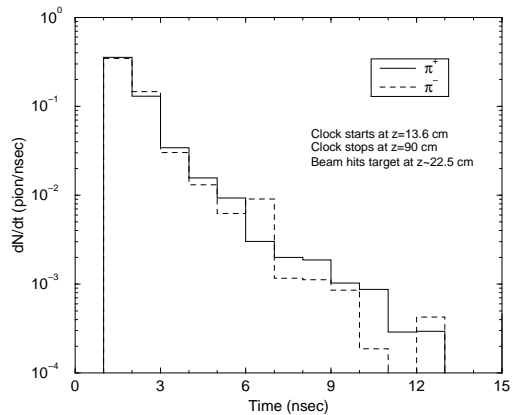, width=3.0in}
\caption{Pion time spectra at L=90~cm and R$<$7.5~cm
for 0.05$<$p$<$0.8~GeV/c and a tilt angle $\alpha$=150~mrad.} 
\label{dndt}
\end{figure}

\begin{figure}[hbt]
\centering
\epsfig{file=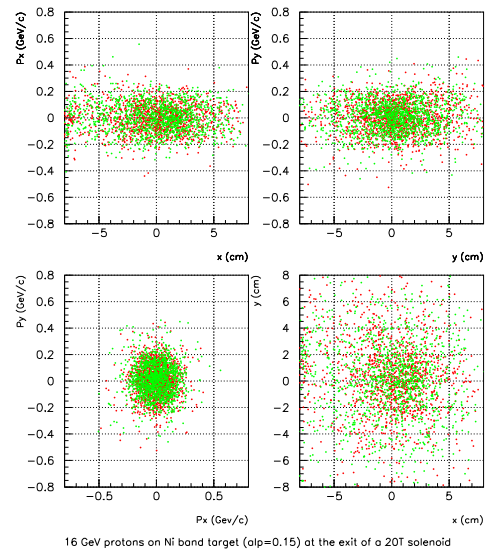, width=3.0in}
\caption{$\pi^+$ (red) and $\pi-$ (green) scatter plots at L=90~cm 
and R$<$7.5~cm for a tilt angle $\alpha$=150~mrad.} 
\label{scatter}
\end{figure}

 The yield per proton at 90 cm downstream from the central
intersection of the beam with the target was determined 
for pions plus muons in the momentum range 0.05$<$p$<$0.8~GeV/c. 
The yields of positive and negative pions were, respectively,
$Y_+$ = 0.491 and $Y_-$ = 0.498 at $\alpha$=0
and  $Y_+$ = 0.622 and $Y_-$ = 0.612 at $\alpha$=150~mrad.
Figure~\ref{phadron} shows the momentum spectra
for all hadrons and figure~\ref{ppion} gives more detailed
information for the pions. Figure~\ref{dndt} shows the
time distribution for when these pions are formed and
figure~\ref{scatter} shows several scatter plots to illustrate
their distribution in phase space.
These pion yields and densities in phase space are comparably good to
the predictions for the best of the liquid jet targets under consideration.

 The peak energy deposition density was found to be 68.6~J/g per
pulse, corresponding to a temperature rise of $\Delta T$=150.5$^{\circ}$C.
This corresponded to a total power dissipation in the target of 0.324~MW
at $\alpha$=150~mrad. Contributions to the deposited energy come from
dE/dx from hadrons and muons (44\%), electromagnetic showering (46\%) and
from absorbtion of sub-threshold particles (10\%).
Power dissipation in inner layer of tungsten shielding (7.5$<$r$<$15~cm)
was also determined, and was found to be
0.624~MW at $\alpha$=0 and 0.766~MW at $\alpha$=150~mrad.

\begin{figure}[hbt]
\centering
\epsfig{file=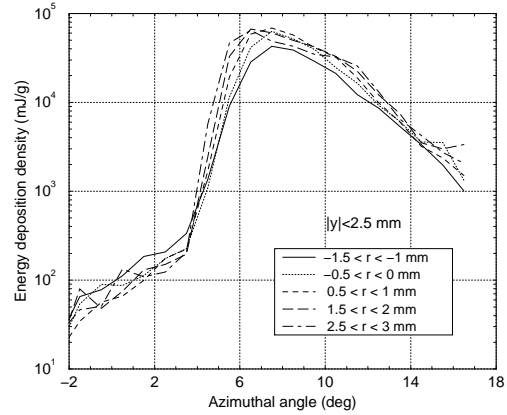, width=3.0in}
\caption{Energy deposition density in the band target \textit{versus}
azimuthal angle for a tilt angle $\alpha$=150~mrad.} 
\label{Edensity}
\end{figure}

\section{ANSYS Stress Simulations}

  The survivability of solid targets in the face of repeated
shock heating is probably the most challenging problem faced
in these scenarios for pion production for muon colliders.

 To investigate this, we are beginning to conduct
finite element computer simulations of the shock heating stresses
using ANSYS, a commercial package that is very widely
used for stress and thermal calculations. These studies are still
at an early stage.

  It is encouraging that the instantaneous energy deposition predicted
by MARS of approximately 70 J/g per proton pulse is much less than
the 500-600 J/g depositions in microsecond timescales that
the Fermilab pbar source nickel target routinely operates
at~\footnote{Editting note: this sentence has been modified from
the original version, which erroneously referred to temperature
rather than energy deposition.}. Further,
if the predicted stresses turn out to be higher than, say, 50\%
of the target's tensile strength then possibilities exist for
redimensioning the target and the proton spot size to reduce the
stress.

\section{Conclusions}
\label{sec-conc}

  In conclusion, initial studies indicate that cupronickel rotating
band targets may well be a viable and attractive option to satisfy
the difficult high power targetry requirements of muon colliders.

\section{references}


[1] The Muon Collider Collaboration, ``Status of Muon Collider Research
and Development and Future Plans'', to be submitted to Phys. Rev. E.

\noindent [2] Alessi {\it et al.}, ``An R \& D Program for Targetry
and Capture at a Muon-Collider Source - A Proposal to the BNL AGS
Division''. Spokesperson Kirk T. McDonald, email:
mcdonald@puphep.princeton.edu . 

\noindent [3] N.V Mokhov, ``The MARS Code System User's Guide,
Version 13 (98)'', FERMILAB-FN-628 (Feb. 1998).

\end{document}